\documentclass[aps,pra,reprint,groupedaddress,showkeys]{revtex4-1}
\usepackage{amsmath,graphicx,hyperref,verbatimbox,array,float,xcolor}
\usepackage{indentfirst}
\usepackage{braket}
\usepackage{diagbox}
\usepackage{amssymb}
\usepackage{bbold}
\usepackage{colortbl}
\usepackage{multirow}
\usepackage{subcaption}
\newcommand{\figref}[1]{Fig.~\ref{#1}}

\renewcommand{\eqref}[1]{Eq.~\ref{#1}}
\newcommand{\tr}{{\mathrm{Tr}}}
\newcommand{\centered}[1]{\begin{tabular}{l} #1 \end{tabular}}

\begin{document}

\title{Quantifying entanglement of two-qubit Werner states}
\author{Artur Czerwinski}
\email{aczerwin@umk.pl}
\affiliation{Institute of Physics, Faculty of Physics, Astronomy and Informatics \\ Nicolaus Copernicus University,
Grudziadzka 5, 87--100 Torun, Poland}

\begin{abstract}
In this article, we introduce a framework for entanglement quantification of photon pairs represented by two-qubit Werner states. The measurement scheme is based on the symmetric informationally complete POVM. To make the framework realistic, we impose the Poisson noise on the measured two-photon coincidences. For various settings, numerical simulations were performed to evaluate the efficiency of the framework.
\end{abstract}
\keywords{Werner states, quantum state tomography, entanglement characterization, concurrence}
\maketitle

\section{Introduction}

Quantum systems can feature non-classical correlations which have become an intrinsic part of quantum physics \cite{Einstein1935}. In particular, entanglement has been a subject of intensive research \cite{Horodecki1996,Horodecki2009}. In general, a multi-particle system is entangled if it cannot be written as a convex combination of product states. For many applications, two-qubit quantum states are considered a key resource \cite{Bennett1992,Erhard2020}.

In the case of photons, entanglement can be detected between various degrees of freedom, for example polarization, spatial or temporal. Polarization-entangled photons have been implemented in quantum information protocols, such as quantum key distribution (QKD) \cite{Ekert1991}, superdense coding \cite{Mattle1996}, quantum teleportation \cite{Bennett1993}, quantum computing \cite{Jozsa2003}, quantum interferometric optical lithography \cite{Boto2000}, etc. There are many ways to generate polarization-entangled photon pairs, like spontaneous parametric down-conversion (SPDC) \cite{White1999} or spontaneous four-wave mixing (SFWM) \cite{Takesue2004}.

Quantum state tomography (QST) is inherent to the development of quantum information theory. Any protocol requires well-characterized quantum states. For this reason, in many applications, the ability to determine an accurate mathematical representation of a physical system plays a central role \cite{Banaszek1999,paris04,Takesue2009,Ikuta2017}. In particular, photonic tomography has attracted much attention due to the vast potential of experiments involving single photons \cite{Altepeter2005}. Therefore, in the present work, we consider state tomography and entanglement quantification of photon pairs that can be described by two-qubit Werner states.

In our framework, we postulate that the source can repeatedly perform the same procedure of preparing photon pairs in an unknown quantum state. Thus, we have access to a relatively large number of identical quantum systems. We can assume that each copy from the ensemble is measured only once. For this reason, the post-measurement state of the system is of little interest, whereas all attention is paid to the probabilities of the respective measurement outcomes. Therefore, we follow the \textit{Positive Operator-Valued Measure} (POVM) formalism \cite{Nielsen2000}.

According to postulates of quantum mechanics, measurements are described by a collection $\{ M_k \}$ of positive semi-definite \textit{measurement operators}, acting on a finite-dimensional Hilbert space $\mathcal{H}$, i.e. $\dim \mathcal{H} =d < \infty$. We assume to operate in the standard basis. The index $k$ refers to the results of measurement that may occur in the experiment. The set of measurement operators $\{ M_k \}$ is called a POVM if
\begin{equation}\label{e2.1}
\sum_k M_k = \mathbb{I}_d,
\end{equation}
where $\mathbb{I}_d$ denotes the identity operator. 

The set of measurement operators $\{ M_k \}$ can be used to calculate the probabilities of the possible measurement outcomes which are obtained according to the Born's rule \cite{Born1955}:
\begin{equation}\label{e2.2}
p (k) = \tr ( M_k \rho ),
\end{equation}
where $\rho$ denotes a density matrix of the system in question. Note that the probabilities have to sum up to one, i.e. $\sum_k p (k) = \sum_k \tr ( M_k \rho ) = 1$, which is equivalent to the condition \eqref{e2.1} since $\tr \rho = 1$.

The goal of QST is to estimate the state by using the results of measurement. In the case of a POVM, if a measurement scheme provides complete knowledge about the state of the system, it is said to be an informationally complete POVM (IC-POVM) \cite{Busch1991,DAriano2004,Flammia2005}. For a given system, there might be various different sets of operators which lead to complete state characterization.

Usually, special attention is paid to a particular case of POVMs which is called a symmetric, informationally complete, positive operator-valued measure (SIC-POVM) \cite{Renes2004}. Originally, SIC-POVMs are constructed from rank-one projectors, but their general properties have also been studied \cite{Rastegin2014}.

Let us assume there is a set of $d^2$ normalized vectors $\ket{\xi_k} \in \mathcal{H}$ such that
\begin{equation}
|\braket{ \xi_i | \xi_j } |^2 = \frac{1}{d+1}\hspace{0.5cm} \text{for} \hspace{0.5cm} i\neq j.
\end{equation}
Then, the set of rank-one projectors $\{ \mathcal{P}_k \}$ defined as
\begin{equation}
\mathcal{P}_k := \frac{1}{d} \ket{\xi_k} \!\bra{\xi_k}\hspace{0.5cm} \text{for} \hspace{0.5cm} k = 1,\dots, d^2
\end{equation}
constitutes a symmetric, informationally complete, positive operator-valued measure (SIC-POVM). The measurement scheme implemented in this work is based on the SIC-POVM for $2-$dimensional Hilbert space. Another common approach to two-qubit tomography involves measurement schemes connected with mutually unbiased bases (MUBs), which can be considered overcomplete \cite{Wootters1989,Durt2010,Nasir2020}.

In Sec.~\ref{methods}, we introduce the framework for state reconstruction and entanglement quantification of photon pairs characterized by two-qubit Werner states. To make the framework realistic, we impose the Poisson noise on the measured photon counts. Then, the framework is tested for different average numbers of photon pairs. In Sec.~\ref{results}, the main results are presented. First, in \ref{correlations}, polarization entanglement analysis is provided by means of two-photon coincidences. Next, in \ref {figuresofmerit}, the figures of merit, selected to quantify the performance of the framework, are displayed on graphs and discussed.

\section{Framework for entanglement quantification}\label{methods}

\subsection{Two-qubit Werner states}

In 1989, R.~Werner introduced mixed quantum states which feature non-classical correlations \cite{Werner1989}. This class of states can be represented by means of the flip operator, $\mathbb{F}$, which acts as $\mathbb{F}(\ket{\psi_1} \otimes \ket{\psi_2}) = \ket{\psi_2} \otimes \ket{\psi_1}$ and takes the form:
\begin{equation}\label{w1}
\mathbb{F} = \sum_{i,j=1}^d \ket{i}\!\bra{j} \otimes \ket{j}\!\bra{i},
\end{equation}
where $\{ \ket{i}\otimes\ket{j}\}$ denotes the standard basis in $\mathcal{H} \otimes \mathcal{H}$. Then, the Werner states can be expressed as
\begin{equation}\label{w2}
\rho_W =  \eta\, \mathbb{F} + \zeta \,\mathbb{I}_{d^2},
\end{equation}
where out of the two parameters only one is independent due to the normalization constraint, i.e. $\tr \rho_W = 1$.

In this work, we are particularly interested in two-qubit Werner states (i.e. $d=2$):
\begin{equation}\label{w3}
\rho_W ^{2q} (\eta) = \eta \ket{\Psi^-}\!\bra{\Psi^-} + \frac{1- \eta}{4} \mathbb{I}_4, 
\end{equation}
where $\ket{\Psi^-} = (\ket{01} - \ket{10})/  \sqrt{2}$ is one of the Bell states and $0\leq \eta \leq 1$. We use $\{\ket{00}, \ket{01}, \ket{10}, \ket{11}\}$ to denote the standard basis in the two-qubit Hilbert space. The states $\rho_W ^{2q} (\eta)$ feature entanglement for any $\eta >1/3$. The class of two-qubit Werner states can be realized by polarization-entangled photon pairs \cite{Barbieri2004}.

Werner states play an important role in quantum information theory, including entanglement purification \cite{Bennett1996a} and quantum teleportation \cite{Lee2000,Yeo2002}. These states have also been used for a description of noisy quantum channels \cite{Checinska2007}. Properties of Werner states in arbitrary dimensions, such as concurrence-based entanglement measures \cite{Chen2006} and nonlocal characteristics based on the Clauser-Horne-Shimony-Holt inequality \cite{Vertesi2008}, remain relevant topics. Therefore, it appears justified to investigate quantum tomography of two-qubit Werner states with noisy measurements.

\subsection{Measurements}

To extract information necessary for entanglement quantification, we utilize a measurement scheme based on the SIC-POVM. When $\dim \mathcal{H} =2$, the SIC-POVM is defined by means of four vectors:
\begin{equation}\label{eqm1}
\begin{split}
&\ket{\phi_1} = \ket{0} \hspace{2.9cm} \ket{\phi_2} = \frac{1}{\sqrt{3}} \ket{0} + \sqrt{\frac{2}{3}} \ket{1} \\
&\ket{\phi_3} = \frac{1}{\sqrt{3}} \ket{0} + \sqrt{\frac{2}{3}} e^{i \frac{2 \pi}{3}} \ket{1}\; \ket{\phi_4} = \frac{1}{\sqrt{3}} \ket{0} + \sqrt{\frac{2}{3}}  e^{i \frac{4 \pi}{3}} \ket{1},
\end{split}
\end{equation}
where $\{\ket{0},\ket{1}\}$ denotes the standard basis in $\mathcal{H}$. Then, the measurement operators are defined as rank$-1$ projectors:
\begin{equation}\label{eqm2}
M_i := \frac{1}{2} \ket{\phi_i}\! \bra{\phi_i},
\end{equation}
which satisfy $\sum_{i=1}^4 M_i = \mathbb{I}_2$. These four measurement operators are sufficient to perform single-qubit tomography \cite{Rehacek2004}. This set of measurements is minimal, but at the same time it allows for efficient and reliable single-qubit tomography. SIC-POVMs, which belong to the class of minimal informationally complete quantum measurements, are in many ways optimal since the ideal measurements in quantum physics are not orthogonal bases \cite{DeBrota2020}.

The measurement scheme can be implemented with the present technology on photons, by a conceptually simple experimental setup, to characterize the polarization state of light\cite{Rehacek2004}. The four-output arrangement provides optimal complete tomography with the minimal number of output channels. It is particularly well-suited as a detection device for quantum communication protocols, such as tomographic quantum cryptography. The SIC-POVM offers a more efficient and very robust approach to QKD than other tomographic protocols \cite{Renes2004a}.

As for two-qubit Werner states, we assume that the source can produce photons pairs that can be characterized by \eqref{w3}. Each photon travels in a separate arm of the experimental setup and is measured individually. Therefore, to determine the quantum state of a pair of photons, we introduce two-qubit measurement operators:
\begin{equation}\label{eqm3}
M^{2q}_{\alpha} := M_i \otimes M_j,
\end{equation}
where $i,j =1, \dots, 4$ and for simplicity we denoted the two-qubit operators with one index, i.e. $\alpha \equiv (i,j)$. From the definition \eqref{eqm3} we see that there are $16$ two-qubit measurement operators.

\subsection{Methods of state reconstruction}

We investigate the efficiency of the measurement scheme based on the SIC-POVM under conditions that include the Poisson noise, which is a typical source of error in single-photon counting \cite{Hernandez2007,Hasinoff2014,Shin2015}. In the two-qubit framework, we assume that the source provides $\mathcal{N}$ polarization-entangled photon pairs per measurement. Each photon travels in a separate arm of the setup and undergoes a measurement described by one of the operators from the SIC-POVM. Then, the detectors receive coincidence counts, $n^{2q}_{\alpha}$, which can be modeled numerically as:
\begin{equation}\label{eqm5}
n^{2q}_{\alpha} = \widetilde{\mathcal{N}}_{\alpha} \,\tr \left(M^{2q}_{\alpha} \, \rho_W^{2q} (\eta)  \right),
\end{equation}
where $\widetilde{\mathcal{N}}_{\alpha} \in \mathrm{Pois} (\mathcal{N})$, i.e. the number of photon pairs for each act of measurement is selected randomly from the Poisson distribution characterized by the mean value $\mathcal{N}$. Therefore, the measured counts are statistically independent Poissonian random variables \cite{Thew2002}. Based on \eqref{eqm5} we can numerically generate noisy data for any two-qubit Werner state $\rho_W^{2q} (\eta)$.

However, when we perform QST, we assume to know nothing about the state in question. For this reason, the expected photon counts are given as
\begin{equation}\label{eqm6}
c^{2q}_{\alpha} = \mathcal{N} \,\tr \left(M^{2q}_{\alpha} \, \sigma^{2q} \right),
\end{equation}
where $\sigma^{2q}$ represents a general $4 \times 4$ density matrix which can be factorized according to the Cholesky decomposition:
\begin{equation}\label{eqm7}
\sigma^{2q} = \frac{T^{\dagger} T}{ \tr (T^{\dagger} T)}
\end{equation}
and $T$ stands for a lower-triangular matrix which depends on $16$ real parameters: $t_1, \dots, t_{16}$. The Cholesky decomposition is commonly implemented in QST frameworks since it guarantees that the result of estimation is physical, i.e. $\sigma^{2q}$ is Hermitian, positive semi-defnite, of trace one \cite{James2001,dariano03}.

The problem of quantum state estimation reduces to determining the parameters which characterize $T$. To find out the values of the parameters which optimally fit to the noisy measurements, we implement the $\chi^2-$estimation, see, e.g., Ref.~\cite{Jack2009}. Thus, we search for the minimum value of the following function:
\begin{equation}\label{eqm8}
\chi^2 (t_1, \dots, t_{16}) = \sum_{\alpha=1}^{16} \frac{(n^{2q}_{\alpha} -  c^{2q}_{\alpha})^2}{c^{2q}_{\alpha}}.
\end{equation}

This procedure allows one to simulate an experimental scenario for any input state $\rho_W^{2q} (\eta)$ -- first we generate noisy photon counts \eqref{eqm5} and then we can recover the state by finding the parameters $t_1, \dots, t_{16}$ for which the $\chi^2$ function reaches its minimum.

\subsection{Performance analysis}

Three figures of merit are introduced to investigate the performance of the measurement scheme on two-qubit Werner states. First, every input Werner state  $\rho_W^{2q} (\eta)$ is compared with the result of $\chi^2-$estimation, $\sigma^{2q}$, by computing the quantum fidelity \cite{Uhlmann1986,Jozsa1994,Bengtsson2006}:
\begin{equation}\label{eqm9}
\mathcal{F}  (\sigma^{2q}, \rho_W^{2q}  (\eta)) := \left(\tr \sqrt{\sqrt{\sigma^{2q}} \, \rho_W^{2q} (\eta)  \, \sqrt{\sigma^{2q}}} \right)^2.
\end{equation}

This figure is commonly used to assess the accuracy of QST frameworks, in particular, under imperfect measurement settings, see, e.g., Ref.~\cite{Rosset2012,Yuan2016,Titchener2018}. In our scenario, the quantum fidelity is treated as a function of $\eta$ to track the precision of the framework along the domain of Werner states.

Next, we analyze how much the states are mixed. Thus, we follow the standard formula for the purity \cite{Nielsen2000}, which implies that we compute $\gamma \equiv \tr (  \rho^2 )$, where $\rho$ denotes a density matrix. The purity of the states resulting from the QST framework $\sigma^{2q}$ is calculated and compared with the input states $\rho_W^{2q} (\eta)$.

Since the main goal of this work is to quantify the amount of entanglement that can be determined through the noisy measurement scheme based on the SIC-POVM, we implement an entanglement measure. For two-qubit Werner states, we compute the concurrence, $C[\rho]$, which quantifies the amount of entanglement in the system described by the density matrix $\rho$ \cite{Hill1997,Wootters1998}. This figure is directly related to another measure, i.e., entanglement of formation \cite{Bennett1996}, but the concurrence can be used as an independent indicator because it is an entanglement monotone. Since there is a straightforward formula to calculate concurrence for any $4 \times 4$ density matrix, it is commonly applied to quantify the amount of entanglement measured by imperfect measurement schemes, see, e.g., Ref.~\cite{Walborn2006,Neves2007,Bergschneider2019}. For two-qubit Werner states, $\rho_W^{2q} (\eta)$, the concurrence is a linear function of $\eta$ \cite{Chen2006}. This theoretical value can be compared with the concurrence of the corresponding estimates $\sigma^{2q}$ in different measurement settings.

\section{Results and analysis}\label{results}

We consider three measurement scenarios that differ in the number of photon pairs provided by the source per measurement. To be more specific, we assume that $\mathcal{N} = 10$, $\mathcal{N} =100$, and $\mathcal{N} = 1\,000$. This allows us to investigate the performance of the framework versus the amount of Poisson noise which is strictly connected with the average number of systems involved in one measurement.

\subsection{Polarization entanglement analysis}\label{correlations}

\begin{figure}[h]
	\centering
  \begin{subfigure}{1\columnwidth}
\caption{$\mathcal{N}=1\,000$}
       \centering
         \includegraphics[width=0.95\columnwidth]{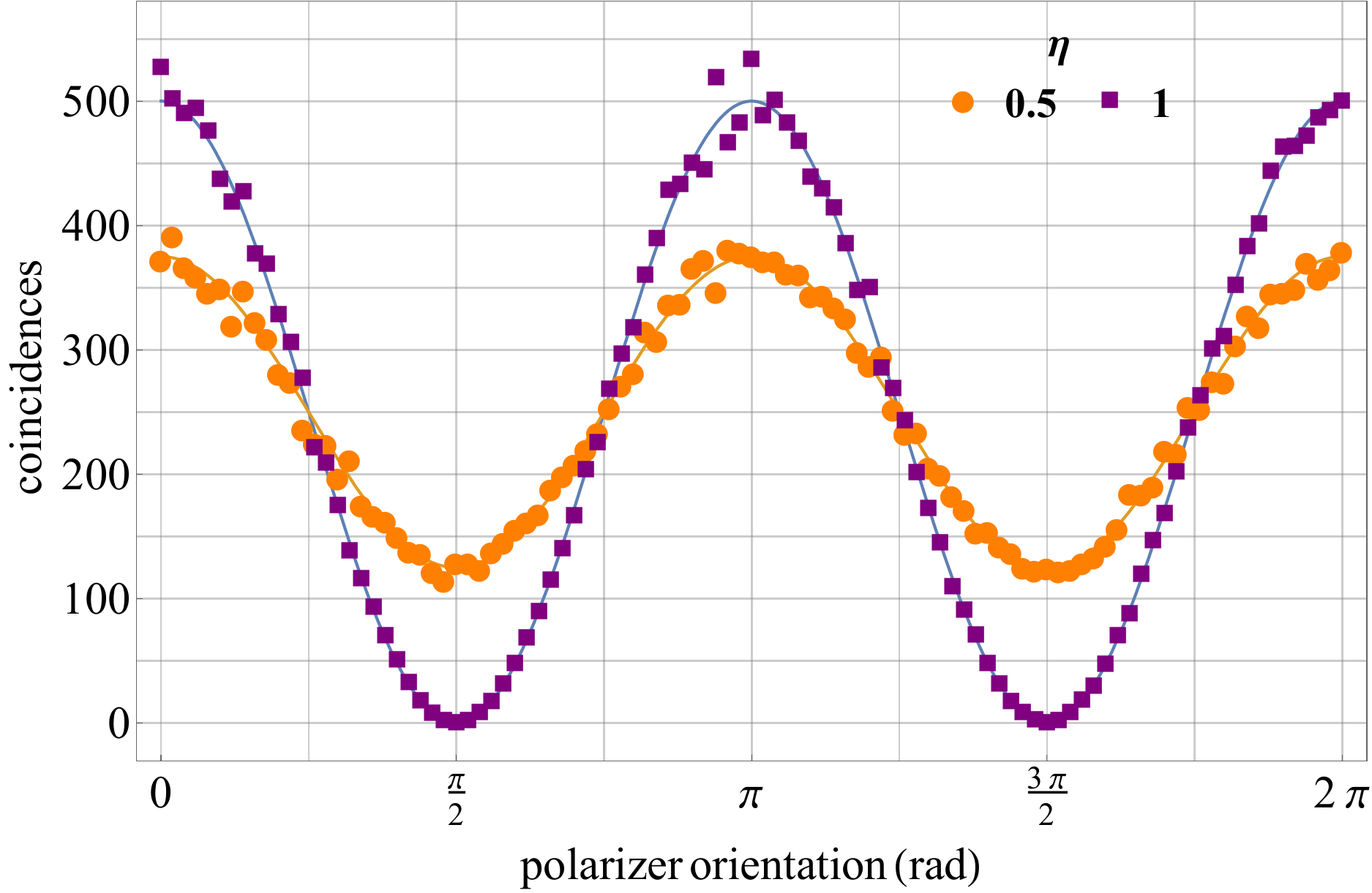}
\label{analysis1}
     \end{subfigure}
     \hfill
     \begin{subfigure}{1\columnwidth}
\caption{$\mathcal{N}=100$}
       \centering
         \includegraphics[width=0.95\columnwidth]{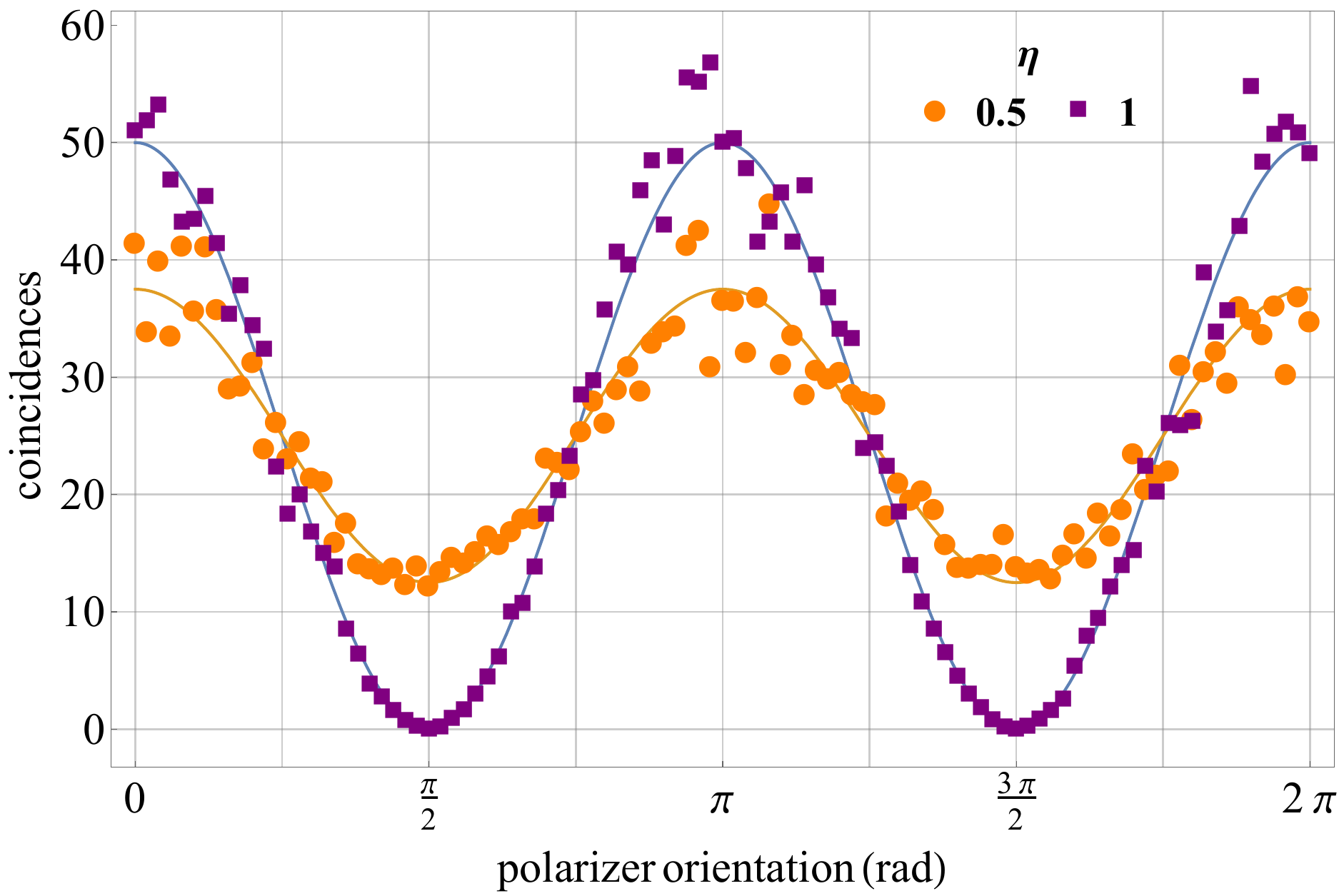}
\label{analysis2}
     \end{subfigure}
     \hfill
     \begin{subfigure}{1\columnwidth}
\caption{$\mathcal{N}=10$}
       \centering
         \includegraphics[width=0.95\columnwidth]{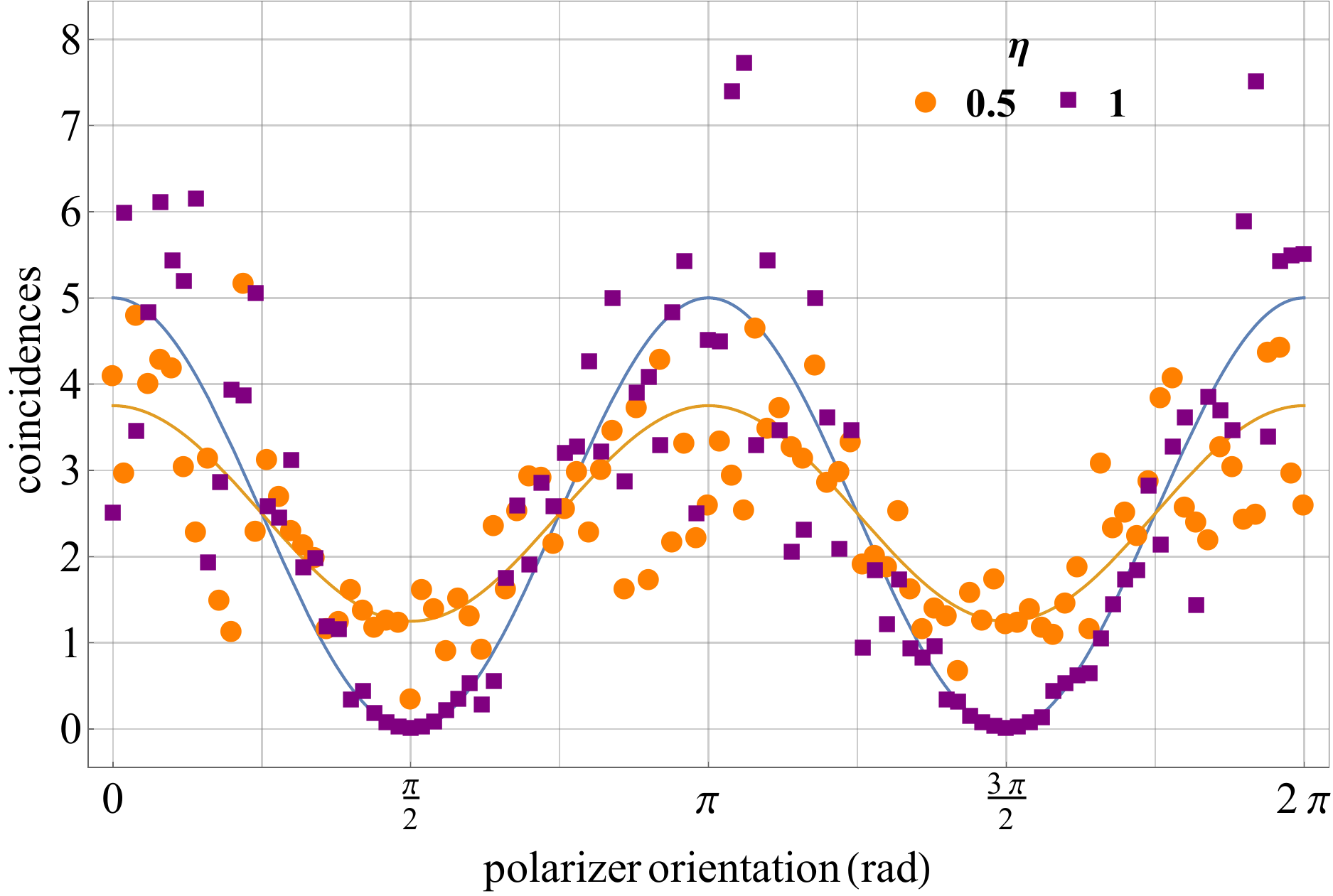}
\label{analysis3}
     \end{subfigure}
	\caption{Entanglement analysis by simulated polarization measurements. The polarizer in one arm was fixed at horizontal orientation while the other made many linear polarization measurements.}
	\label{plots1}
\end{figure}

First, the simulations were performed to display quantum correlations of two-qubit Werner states. We consider two specific cases, i.e., $\eta=0.5$ and $\eta=1$, such that the states of the form \eqref{w3} feature entanglement. The measurement scheme was realized by assuming that the polarizer in one arm is fixed at horizontal (H) orientation whereas the other polarizer rotates to make many linear polarization measurements (zero angle corresponds to the vertical (V) orientation in the other polarizer). This technique allows one to count two-photon coincidences which indicate non-classical correlations of the quantum system, see, e.g., Ref.~\cite{Kwiat1999,Horn2013,Nomerotski2020}.  In \figref{plots1}, three series of average two-photon counts are presented, assuming different numbers of photon pairs involved in a single measurement.

In \figref{plots1}, the continuous lines present expected measurement results according to the theoretical model without noise. The simulated measurement results burdened with the Poisson noise are given by dots (for $\eta=0.5$) or by squares (for $\eta=1$). For different numbers of photon pairs per measurement, the plots demonstrate how the two-photon coincidences deviate from the expected course. In particular, in \figref{analysis3}, we notice a significant amount of deviation. These results confirm the hypothesis that the Poisson noise is more detrimental if we utilize fewer photon pairs per measurement.

\subsection{Quantifying entanglement}\label{figuresofmerit}

\begin{figure}[h]
	\centering
         \centered{\includegraphics[width=0.95\columnwidth]{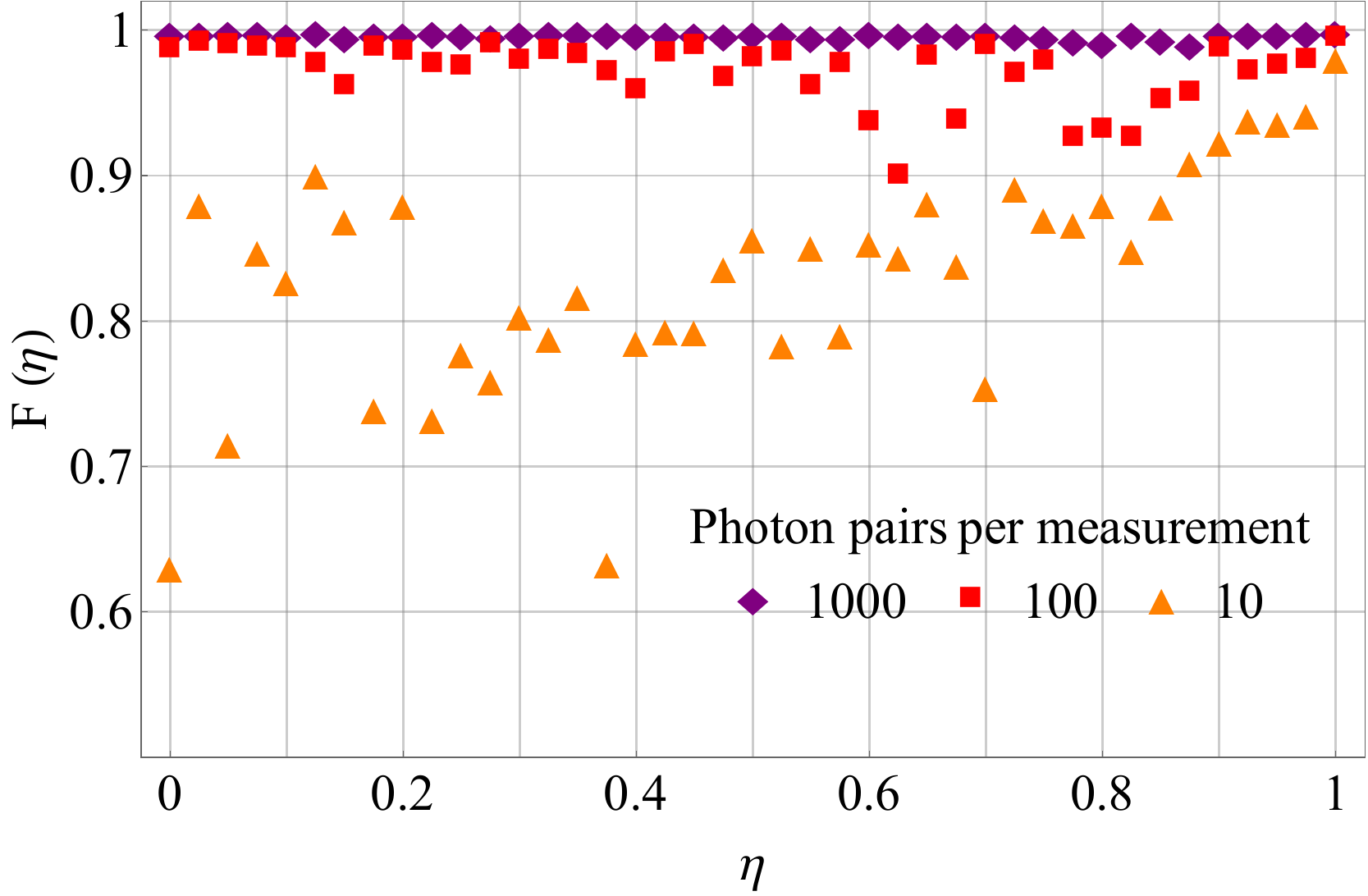}}
	\caption{Plots present the quantum fidelity, $\mathcal{F}  (\eta)$, in QST of two-qubit Werner states for three different numbers of photon pairs per measurement.}
	\label{fidelity1}
\end{figure}

For two-qubit Werner states, we first evaluate the accuracy of state reconstruction by quantum fidelity. In \figref{fidelity1}, one can observe the plots for three different numbers of photon pairs produced by the source per measurement. The simulations were performed along the entire domain of $\eta$.

The results demonstrate that $\mathcal{N}=1\,000$ is sufficient for precise state reconstruction since the plot corresponding to this number resembles a constant function with value $\approx 1$. Then, we notice that if the number of photon pairs is decreased to $\mathcal{N}=100$, we obtain some deviations from the desirable value of quantum fidelity. However, it is worth noting that in this case $\mathcal{F} (\eta) > 0.9$ for all $\eta$.

Furthermore, when we reduce the number of photon pairs to $\mathcal{N}=10$, we notice a detrimental impact of the Poisson noise on the accuracy of state reconstruction. We observe that the function of quantum fidelity is irregular, which can be attributed to the randomness of noise. In this scenario, we have $\mathcal{F} (\eta) > 0.6$ for all $\eta$. Nonetheless, the quantum fidelity improves as we increase $\eta$. In fact, for $\eta=1$ all three plots converge. This suggests that the single-photon scenario is efficient in QST of pure Werner states.

\begin{figure}[h]
	\centering
         \centered{\includegraphics[width=0.95\columnwidth]{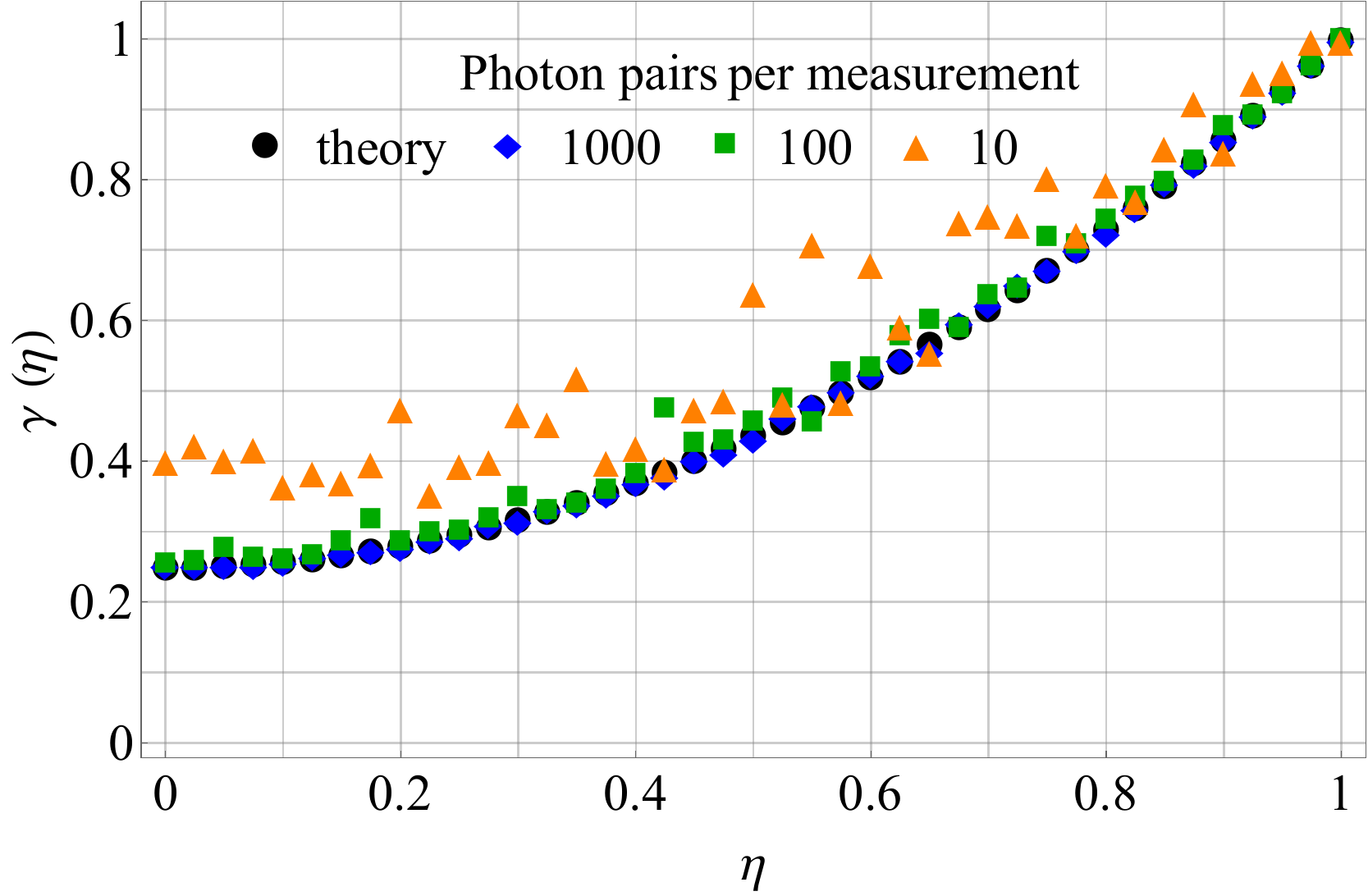}}
	\caption{Plots present the purity, $\gamma(\eta)$, in QST of two-qubit Werner states for three different numbers of photon pairs per measurement.}
	\label{purity1}
\end{figure}

In the next step, we study the purity of reconstructed states and compare it with the purity for the actual two-qubit Werner states. In \figref{purity1}, one can observe that the plot for $\mathcal{N}=1\,000$ overlaps with the theoretical value, whereas for $\mathcal{N}=100$ we witness minor deteriorations. For $\mathcal{N}=10$, we note significant discrepancies between the estimated and the actual purity. Interestingly, in most cases, the Poisson noise distorts the purity in such a way that it is greater than the actual value.

\begin{figure}[h]
	\centering
         \centered{\includegraphics[width=0.95\columnwidth]{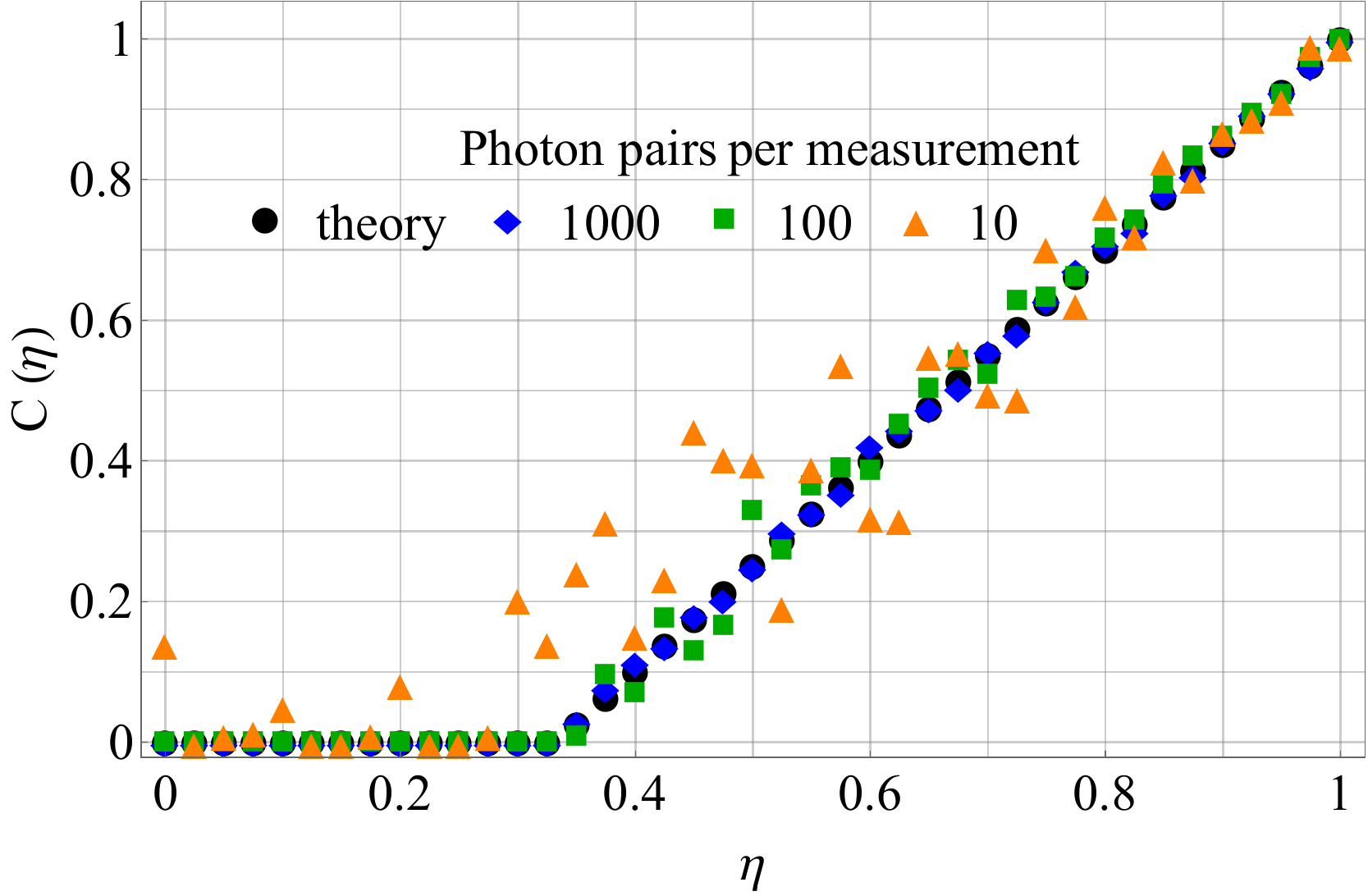}}
	\caption{Plots present the concurrence, $ C (\eta)$, in QST of two-qubit Werner states for three different numbers of photon pairs per measurement.}
	\label{concurrence1}
\end{figure}

Finally, we examine entanglement quantification through noisy measurements. In \figref{concurrence1}, we observe the concurrence of the estimates for three numbers of photon pairs. The results can be compared with the theoretical value of concurrence of the two-qubit Werner states. Again, we discover a perfect consistency between the theoretical value and the results for $\mathcal{N}=1\,000$. For the middle number of photon pairs, we identify minor inaccuracies. For $\mathcal{N}=10$, the concurrence of the estimated states departs significantly from the actual value. It is worth noting that often the estimated concurrence is greater than the actual value. This might lead to a misleading impression that the single-photon scenario performs better at entanglement quantification, which is not the case since we cannot measure more entanglement than the input states feature. Especially, if we consider $\eta \leq 1/3$, the input states $\rho_W^{2q} (\eta)$ are separable, whereas some estimates feature entanglement, which should be understood as an error due to the substantial impact of the Poisson noise in the single-photon scenario.

To conclude, we have found out that all the plots in \figref{fidelity1}-\ref{concurrence1} converge, which means that for pure (or almost pure) Werner states, the QST framework is equally efficient, irrespective of the number of photon pairs involved in a single measurement. This result appears unanticipated since in \figref{plots1} it was demonstrated that the measurements for $\mathcal{N}=10$ feature the greatest amount of noise.Nevertheless, from a practical point of view, these findings imply that we can efficiently implement the framework for the entire class of two-qubit Werner states provided we utilize many photon pairs per measurement. With single photons at our disposal, we can accurately measure only those Werner states that are close to pure.

\section{Discussion and summary}\label{discussion}

In the article, we introduced a framework for entanglement quantification of two-qubit Werner states that are encoded on photon pairs in the polarization degree of freedom. The efficiency of the framework was investigated for three average numbers of photon pairs: $10,\,10^2,\,10^3$. The results have revealed that if we utilize $10^3$ photon pairs per measurement, the framework is robust against the Poisson noise. On this assumption, the two-photon coincidences and concurrence precisely fit the theoretical expectations. Additionally, the quantum fidelity is, approximately, a constant function with the value equal one, which ensures high-accuracy state tomography along the entire domain.

If the average number of photon pairs was reduced to $10$, we observed severe inaccuracies in the two-photon coincidences. Furthermore, in the single-photon scenario, the framework appeared unreliable at quantifying entanglement of mixed two-qubit Werner states. We discovered a deceptive effect which suggested that the single-photon scenario could have measured more entanglement, which was only a result of the Poisson noise. For some input states, the concurrence of the reconstructed density matrix exceeded the theoretical value corresponding to the Werner state. Similarly, in the case of purity, the estimated states featured less entropy than expected. Both effects demonstrate the impact of the Poisson noise on entanglement detection and state characterization.

Furthermore, it was found out that all figures of merit converge at $\eta =1$, which implies that for almost pure two-qubit Werner states, the framework is equally efficient, irrespective of the number of photon pairs involved in each measurement. Therefore, one can state that the single-photon scenario is advisable only if we are certain that the state in question is close to pure.

In the future, the framework presented in the article will be generalized to higher-dimensional Werner states. There are two significant challenges that need to be tackled to achieve this goal. More specifically, the first limitation concerns the existence of SIC-POVM in an arbitrary Hilbert space. As we know, it remains an open problem and, for this reason, there might be a need to apply other measurement schemes for multilevel entangled states. Secondly, the concurrence, which was used in the present manuscript to quantify entanglement, cannot be generalized in a straightforward way. For numerical simulations, it is necessary to operate with an entanglement measure that is computable for any estimate obtained from the QST framework. Therefore, a part of future research will involve defining an appropriate measure based, for example, on a geometric distance, see Ref.~\cite{Vedral1997}. Due to the complexity of these problems, such analysis will be conducted in forthcoming articles.

\section*{Acknowledgments}

The author acknowledges financial support from the Foundation for Polish Science (FNP) (project First Team co-financed by the European Union under the European Regional Development Fund).


\begin{thebibliography}{1}

\bibitem{Einstein1935}
A.~Einstein, B.~Podolsky, and N.~Rosen, \textit{Phys. Rev.} \textbf{47}, 777-780 (1935) \doi{10.1103/PhysRev.47.777}

\bibitem{Horodecki1996}
M.~Horodecki, P.~Horodecki, and R.~Horodecki, \textit{Phys. Lett. A} \textbf{223}, 1-8 (1996) \doi{10.1016/S0375-9601(96)00706-2}

\bibitem{Horodecki2009}
R.~Horodecki, P.~Horodecki, M.~Horodecki, and K.~Horodecki, \textit{Rev. Mod. Phys.} \textbf{81}, 865-942 (2009) \doi{10.1103/RevModPhys.81.865}

\bibitem{Bennett1992}
C.~H.~Bennett and S.~J.~Wiesner, \textit{Phys. Rev. Lett.} \textbf{69}, 2881-2884 (1992) \doi{10.1103/PhysRevLett.69.2881}

\bibitem{Erhard2020}
M.~Erhard, M.~Krenn, and A.~Zeilinger, \textit{Nat. Rev. Phys.} \textbf{2}, 365-381 (2020) \doi{10.1038/s42254-020-0193-5}

\bibitem{Ekert1991}
A.~K.~Ekert, \textit{Phys. Rev. Lett.} \textbf{67}, 661-663 (1991) \doi{10.1103/PhysRevLett.67.661}

\bibitem{Mattle1996}
K.~Mattle, H.~Weinfurter, P.~G.~Kwiat, and A.~Zeilinger, \textit{Phys. Rev. Lett.} \textbf{76}, 4656-4659 (1996) \doi{10.1103/PhysRevLett.76.4656}

\bibitem{Bennett1993}
C.~H.~Bennett, G.~Brassard, C.~Crepeau, R.~Jozsa, A.~Peres, and W.~K.~Wootters, \textit{Phys. Rev. Lett.} \textbf{70}, 1895-1899 (1993) \doi{10.1103/PhysRevLett.70.1895}

\bibitem{Jozsa2003}
R.~Jozsa and N.~Linden, \textit{Proc. R. Soc. A} \textbf{459}, 2011-2032 (2003) \doi{10.1098/rspa.2002.1097}

\bibitem{Boto2000}
A.~N.~Boto, P.~Kok, D.~S.~Abrams, S.~L.~Braunstein, C.~P.~Williams, and J.~P.~Dowling, \textit{Phys. Rev. Lett.} \textbf{85}, 2733-2736 (2000) \doi{10.1103/PhysRevLett.85.2733}

\bibitem{White1999}
A.~G.~White, D.~F.~V.~James, P.~H.~Eberhard, and P.~G.~Kwiat, \textit{Phys. Rev. Lett.} \textbf{83}, 3103-3106 (1999) \doi{10.1103/PhysRevLett.83.3103}

\bibitem{Takesue2004}
H.~Takesue and K.~Inoue, \textit{Phys. Rev. A} \textbf{70}, 031802(R) (2004) \doi{10.1103/PhysRevA.70.031802}

\bibitem{Banaszek1999}
K.~Banaszek, G.~M.~D'Ariano, M.~G.~A.~Paris, and M.~F.~Sacchi, \textit{Phys. Rev. A} \textbf{61}, 010304(R) (1999) \doi{10.1103/PhysRevA.61.010304}

\bibitem{paris04}
M.~G.~A.~Paris and J.~\v{R}eh\'{a}\v{c}ek (eds.), \textit{Quantum State Estimation (Lecture Notes in Physics)}, Springer, Berlin-Heidelberg (2004) \doi{10.1007/b98673}

\bibitem{Takesue2009}
H.~Takesue and Y.~Noguchi, \textit{Opt. Express.} \textbf{17}, 10976 (2009) \doi{10.1364/OE.17.010976}

\bibitem{Ikuta2017}
T.~Ikuta and H.~Takesue, \textit{New J. Phys.} \textbf{19}, 013039 (2017) \doi{10.1088/1367-2630/aa5571}

\bibitem{Altepeter2005}
J.~Altepeter, E.~Jerey, and P.~Kwiat, \textit{Adv. At. Mol. Opt. Phys.} \textbf{52}, 105-159 (2005) \doi{10.1016/S1049-250X(05)52003-2}

\bibitem{Nielsen2000}
M.~A.~Nielsen and I.~L.~Chuang, \textit{Quantum Computation and Quantum Information}, Cambridge University Press, Cambridge (2000) \doi{10.1017/CBO9780511976667}

\bibitem{Born1955}
M.~Born, \textit{Science} \textbf{122}, 675-679 (1955) \doi{10.1126/science.122.3172.675}

\bibitem{Busch1991}
P.~Busch, \textit{Int. J. Theor. Phys.} \textbf{30}, 1217-1227 (1991) \doi{10.1007/BF00671008}

\bibitem{DAriano2004}
G.~M.~D'Ariano, P.~Perinotti, M.~F.~Sacchi, \textit{J. Opt. B: Quantum Semiclass. Opt.} \textbf{6}, 487-491 (2004) \doi{10.1088/1464-4266/6/6/005}

\bibitem{Flammia2005}
S.~T.~Flammia, A.~Silberfarb, C.~M.~Caves, \textit{Found. Phys.} \textbf{35}, 1985-2006 (2005) \doi{10.1007/s10701-005-8658-z}

\bibitem{DAriano2000}
G.~M.~D'Ariano, L.~Maccone, M.~Paris, \textit{Phys. Lett. A} \textbf{276}, 25-30 (2000) \doi{10.1016/S0375-9601(00)00660-5}

\bibitem{Renes2004}
J.~M.~Renes, R.~Blume-Kohout, A.~J.~Scott, C.~M.~Caves, \textit{J. Math. Phys.} \textbf{45}, 2171-2180 (2004) \doi{10.1063/1.1737053}

\bibitem{Rastegin2014}
A.~E.~Rastegin, \textit{Phys. Scr.} \textbf{89}, 085101 (2014) \doi{10.1088/0031-8949/89/8/085101}

\bibitem{Wootters1989}
W.~K.~Wootters and B.~D.~Fields, \textit{Ann. Phys.} \textbf{191}, 363-381 (1989)  \doi{10.1016/0003-4916(89)90322-9}

\bibitem{Durt2010}
T.~Durt, B.-G.~Englert, I.~Bengtsson, and K.~\.Zyczkowski, \textit{Int. J. Quantum Inf.} \textbf{8}, 535-640 (2010) \doi{10.1142/S0219749910006502}

\bibitem{Nasir2020}
R.~N.~M.~Nasir, J.~S.~Shaari, and S.~Mancini, \textit{Int. J. Quantum Inf.} \textbf{18}, 1941026 (2020) \doi{10.1142/S0219749919410260}

\bibitem{Werner1989}
R.~F.~Werner, \textit{Phys. Rev. A} \textbf{40}, 4277-4281 (1989) \doi{10.1103/PhysRevA.40.4277}

\bibitem{Barbieri2004}
M.~Barbieri, F.~De~Martini, G.~Di~Nepi, and P.~Mataloni, \textit{Phys. Rev. Lett.} \textbf{92}, 177901 (2004) \doi{10.1103/PhysRevLett.92.177901}

\bibitem{Bennett1996a}
C.~H.~Bennett, G.~Brassard, S.~Popescu, B.~Schumacher, J.~A.~Smolin, and W.~K.~Wootters, \textit{Phys. Rev. Lett.} \textbf{76}, 722 (1996) \doi{10.1103/PhysRevLett.76.722}

\bibitem{Lee2000}
J.~Lee and M.~S.~Kim, \textit{Phys. Rev. Lett.} \textbf{84}, 4236 (2000) \doi{10.1103/PhysRevLett.84.4236} 

\bibitem{Yeo2002}
Y.~Yeo, Phys. Rev. A \textbf{66}, 062312 (2002) \doi{10.1103/PhysRevA.66.062312}

\bibitem{Checinska2007}
A.~Checinska and K. Wodkiewicz, \textit{Phys. Rev. A} \textbf{76}, 052306 (2007) \doi{10.1103/PhysRevA.76.052306}

\bibitem{Chen2006}
K.~Chen, S.~Albeverio, S.-M.~Fei, \textit{Rep. Math. Phys.} \textbf{58}, 325--334 (2006)  \doi{10.1016/S0034-4877(07)00003-1}

\bibitem{Vertesi2008}
T.~Vertesi, \textit{Phys. Rev. A} \textbf{78}, 032112 (2008) \doi{10.1103/PhysRevA.78.032112}

\bibitem{Rehacek2004}
 J.~\v{R}eh\'{a}\v{c}ek, B.-G.~Englert, and D.~Kaszlikowski, \textit{Phys. Rev. A} \textbf{70}, 052321 (2004) \doi{10.1103/PhysRevA.70.052321}

\bibitem{DeBrota2020}
J. B. DeBrota, C. A. Fuchs, and B. C. Stacey, \textit{Int. J. Quantum Inf.} (2020) \doi{10.1142/S0219749920400055}

\bibitem{Renes2004a}
J.~M.~Renes, \textit{Phys. Rev. A} \textbf{70}, 052314 (2004) \doi{10.1103/PhysRevA.70.052314}

\bibitem{Hernandez2007}
S.~Hernandez-Marin, A.~M.~Wallace, and G.~J.~Gibson, \textit{IEEE Trans. Pattern Anal. Mach. Intell.} \textbf{29}, 2170-2180 (2007) \doi{10.1109/TPAMI.2007.1122}

\bibitem{Hasinoff2014}
S.~W.~Hasinoff, \textit{Photon, poisson noise}, in: K.~Ikeuchi (eds.), \textit{Computer Vision}, Springer, Boston, MA, pp. 608-610 (2014) \doi{10.1007/978-0-387-31439-6_482}

\bibitem{Shin2015}
D.~Shin, A.~Kirmani, V.~K.~Goyal and J.~H.~Shapiro, \textit{IEEE Trans. Comput. Imaging.} \textbf{1}, 112-125 (2015) \doi{10.1109/TCI.2015.2453093}

\bibitem{Thew2002}
R.~T.~Thew, K.~Nemoto, A.~G.~White, W.~J.~Munro, \textit{Phys. Rev. A} \textbf{66}, 012303 (2002) \doi{10.1103/PhysRevA.66.012303}

\bibitem{James2001}
D.~F.~V.~James, P.~G.~Kwiat, W.~J.~Munro, and A.~G.~White, \textit{Phys. Rev. A} \textbf{64}, 052312 (2001)  \doi{10.1103/PhysRevA.64.052312}

\bibitem{dariano03}
G.~M.~D'Ariano, M.~G.~A.~Paris, and M.~F.~Sacchi, \textit{Adv. Imaging Electron Phys.} \textbf{128}, 205-308 (2003) \doi{10.1016/S1076-5670(03)80065-4}

\bibitem{Jack2009}
B.~Jack, J.~Leach, H.~Ritsch, S.~M.~Barnett, M.~J.~Padgett, and S.~Franke-Arnold, \textit{New J. Phys.} \textbf{11}, 103024 (2009) \doi{10.1088/1367-2630/11/10/103024}

\bibitem{Uhlmann1986}
A.~Uhlmann, \textit{Rep. Math. Phys.} \textbf{24}, 229-240 (1986) \doi{10.1016/0034-4877(86)90055-8}

\bibitem{Jozsa1994}
R.~Jozsa, \textit{J. Mod. Opt.} \textbf{41}, 2315-2323 (1994) \doi{10.1080/09500349414552171}

\bibitem{Bengtsson2006}
I.~Bengtsson and K.~\.Zyczkowski, \textit{Geometry of Quantum States. An Introduction to Quantum Entanglement}, Cambridge University Press, Cambridge (2006) \doi{10.1017/CBO9780511535048}

\bibitem{Titchener2018}
J.~G.~Titchener, M.~Gr\"afe, R.~Heilmann, A.~S.~Solntsev, A.~Szameit, and A.~A.~Sukhorukov, \textit{npj Quantum Inf.} \textbf{4}, 19 (2018) \doi{10.1038/s41534-018-0063-5}

\bibitem{Yuan2016}
H.~Yuan, Z.-W.~Zhou, and G.-C. Guo, \textit{New J. Phys.} \textbf{18}, 043013 (2016) \doi{10.1088/1367-2630/18/4/043013}

\bibitem{Rosset2012}
D.~Rosset, R.~Ferretti-Sch\"obitz, J.-D.~Bancal, N.~Gisin, and Y.-C.~Liang, \textit{Phys. Rev. A} \textbf{86}, 062325 (2012) \doi{10.1103/PhysRevA.86.062325}

\bibitem{Hill1997}
S.~Hill and W.~K.~Wootters, \textit{Phys. Rev. Lett.} \textbf{78}, 5022-5025 (1997) \doi{10.1103/PhysRevLett.78.5022}

\bibitem{Wootters1998}
W.~K.~Wootters, \textit{Phys. Rev. Lett.} \textbf{80}, 2245-2248 (1998) \doi{10.1103/PhysRevLett.80.2245}

\bibitem{Bennett1996}
C.~H.~Bennett, D.~P.~DiVincenzo, J.~A.~Smolin, and W.~K.~Wootters, \textit{Phys. Rev. A} \textbf{54}, 3824-3851 (1996) \doi{10.1103/PhysRevA.54.3824}

\bibitem{Walborn2006}
S.~P.~Walborn, P.~H.~Souto Ribeiro, L.~Davidovich, F.~Mintert, and A.~Buchleitner, \textit{Nature} \textbf{440}, 1022-1024 (2006) \doi{10.1038/nature04627}

\bibitem{Neves2007}
L.~Neves, G.~Lima, E.~J.~S.~Fonseca, L.~Davidovich, and S.~Padua, \textit{Phys. Rev. A} \textbf{76}, 032314 (2007) \doi{10.1103/PhysRevA.76.032314}

\bibitem{Bergschneider2019}
A.~Bergschneider, V.~M.~Klinkhamer, J.~H.~Becher, R.~Klemt, L.~Palm, G.~Zurn, S.~Jochim, and P.~M.~Preiss, \textit{Nat. Phys.} \textbf{15}, 640-644 (2019) \doi{10.1038/s41567-019-0508-6}

\bibitem{Kwiat1999}
P.~G.~Kwiat, E.~Waks, A.~G.~White, I.~Appelbaum, and P.~H.~Eberhard, \textit{Phys. Rev. A} \textbf{60}, R773(R) (1999) \doi{10.1103/PhysRevA.60.R773}

\bibitem{Horn2013}
R.~T.~Horn, P.~Kolenderski, D.~Kang, P.~Abolghasem, C.~Scarcella, A.~D.~Frera, A.~Tosi, L.~G.~Helt, S.~V.~Zhukovsky, J.~E.~Sipe, G.~Weihs, A.~S.~Helmy, and T.~Jennewein, \textit{Sci. Rep.} \textbf{3}, 2314 (2013)  \doi{10.1038/srep02314}

\bibitem{Nomerotski2020}
A.~Nomerotski, D.~ Katramatos, P.~Stankus, P.~Svihra, G.~Cui, S.~Gera, M.~Flament, and E.~Figueroa, \textit{Int. J. Quantum Inf.} \textbf{18}, 1941027 (2020) \doi{10.1142/S0219749919410272}

\bibitem{Vedral1997}
V.~Vedral, M.~B.~Plenio, M.~A.~Rippin, and P.~L.~Knight, \textit{Phys. Rev. Lett.} \textbf{78}, 2275 (1997) \doi{10.1103/PhysRevLett.78.2275}

\end{thebibliography}
\end{document}